\documentclass[aip,apl,reprint]{revtex4-1} 

\usepackage{graphicx}
\usepackage{dcolumn}
\usepackage{bm}
\usepackage{amsmath, amssymb}
\usepackage{float}
\usepackage{mhchem}

\begin{document}

\title{Free-standing and positionable microwave antenna device for magneto-optical spectroscopy experiments}

\author{T. Hache}
\affiliation{Helmholtz-Zentrum Dresden--Rossendorf, Institute of Ion Beam Physics and Materials Research, Bautzner Landstra\ss e 400, 01328 Dresden, Germany}
\affiliation{Institut f\"ur Physik, Technische Universit\"at Chemnitz, 09107 Chemnitz, Germany}

\author{M. Va\v{n}atka}
\affiliation{CEITEC BUT, Brno University of Technology, Purky\v{n}ova 123, 612 00 Brno, Czech Republic}

\author{L. Flaj\v{s}man}
\affiliation{CEITEC BUT, Brno University of Technology, Purky\v{n}ova 123, 612 00 Brno, Czech Republic}

\author{T. Weinhold}
\affiliation{Helmholtz-Zentrum Dresden--Rossendorf, Institute of Ion Beam Physics and Materials Research, Bautzner Landstra\ss e 400, 01328 Dresden, Germany}
\affiliation{Technische Universit\"at Dresden, 01062 Dresden, Germany}

\author{T. Hula}
\affiliation{Helmholtz-Zentrum Dresden--Rossendorf, Institute of Ion Beam Physics and Materials Research, Bautzner Landstra\ss e 400, 01328 Dresden, Germany}
\affiliation{Institut f\"ur Physik, Technische Universit\"at Chemnitz, 09107 Chemnitz, Germany}

\author{O. Ciubotariu}
\affiliation{Institute of Physics, University of Augsburg, Universit\"atsstra\ss{}e 1, 86135 Augsburg, Germany}

\author{M. Albrecht}
\affiliation{Institute of Physics, University of Augsburg, Universit\"atsstra\ss{}e 1, 86135 Augsburg, Germany}

\author{B. Arkook}
\affiliation{Physics and Astronomy, University of California, Riverside, CA 92521, USA}

\author{I. Barsukov}
\affiliation{Physics and Astronomy, University of California, Riverside, CA 92521, USA}

\author{L. Fallarino}
\affiliation{Helmholtz-Zentrum Dresden--Rossendorf, Institute of Ion Beam Physics and Materials Research, Bautzner Landstra\ss e 400, 01328 Dresden, Germany}

\author{O. Hellwig}
\affiliation{Helmholtz-Zentrum Dresden--Rossendorf, Institute of Ion Beam Physics and Materials Research, Bautzner Landstra\ss e 400, 01328 Dresden, Germany}
\affiliation{Institut f\"ur Physik, Technische Universit\"at Chemnitz, 09107 Chemnitz, Germany}

\author{J. Fassbender}
\affiliation{Helmholtz-Zentrum Dresden--Rossendorf, Institute of Ion Beam Physics and Materials Research, Bautzner Landstra\ss e 400, 01328 Dresden, Germany}
\affiliation{Technische Universit\"at Dresden, 01062 Dresden, Germany}

\author{M. Urb\'{a}nek}
\affiliation{CEITEC BUT, Brno University of Technology, Purky\v{n}ova 123, 612 00 Brno, Czech Republic}

\author{H. Schultheiss}
\affiliation{Helmholtz-Zentrum Dresden--Rossendorf, Institute of Ion Beam Physics and Materials Research, Bautzner Landstra\ss e 400, 01328 Dresden, Germany}
\affiliation{Technische Universit\"at Dresden, 01062 Dresden, Germany}

\date{\today}

\begin{abstract}

Modern spectroscopic techniques for the investigation of magnetization dynamics in micro- and nano-structures or thin films use typically microwave antennas which are directly fabricated on top of the sample by means of electron-beam-lithography (EBL). 
Following this approach, 
every magnetic structure on the sample needs its own antenna, resulting in additional EBL steps and layer deposition processes. 
 We demonstrate a new approach for magnetization excitation that is suitable for optical and non-optical spectroscopy techniques. By patterning the antenna on a separated flexible glass cantilever and insulating it electrically, we solved the before mentioned issues. Since we use flexible transparent glass as a substrate, optical spectroscopy techniques like Brillouin-light-scattering microscopy (${\mu}$BLS), time resolved magneto-optical Kerr effect measurements (TRMOKE) or optical detected magnetic resonance (ODMR) measurements can be carried out at visible laser wavelengths. As the antenna is detached from the sample it can be freely positioned 
in all three dimensions to adress only the desired magnetic sample structures 
and to achieve effective excitation.
We demonstrate the functionality of these antennas using ${\mu}$BLS and 
compare 
coherently and thermally excited magnon spectra to show the enhancement of the signal by a factor of about 400 
due to the excitation by the antenna. Moreover, we succeed to characterize yttrium iron garnet thin films with spatial resolution using optical ferromagnetic resonance (FMR) experiments.
We analyse the spatial excitation profile of the antenna by measuring the magnetization dynamics in two dimensions. 
The technique is furthermore applied to investigate injection-locking of spin Hall nano-oscillators. 
\end{abstract}


\maketitle

\subsection*{Introduction}

\begin{figure}[h]
\begin{center}
\scalebox{1}{\includegraphics[width=7.9 cm, clip]{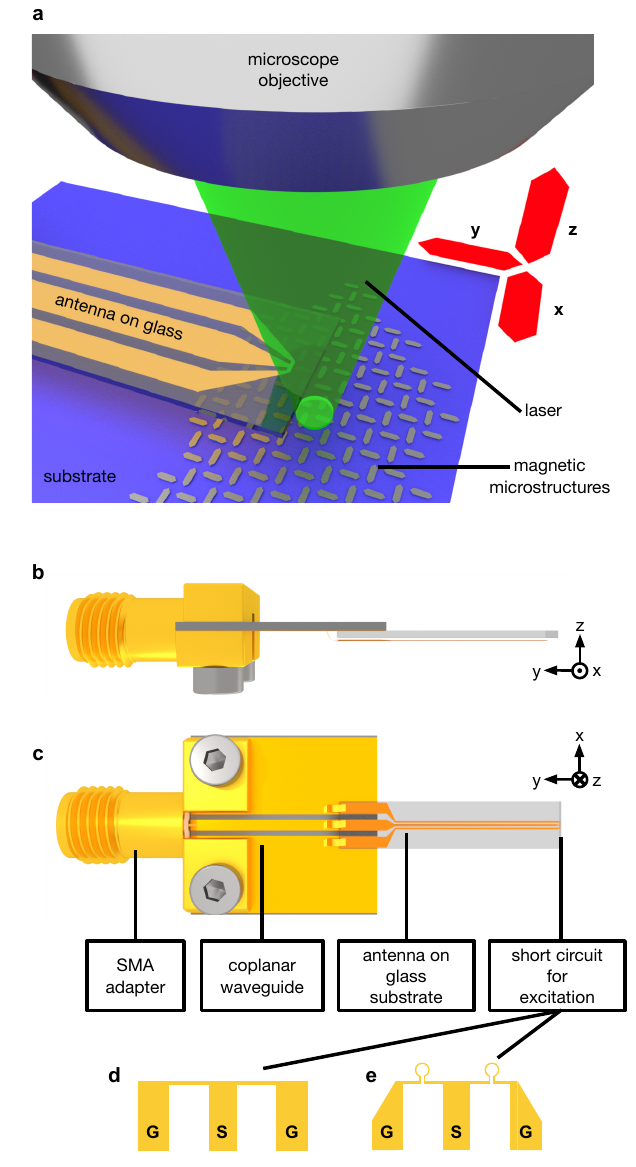}}
\caption{\label{fig7}(color online) (a) Suspended antenna providing local excitation of magnetic elements is fabricated on a glass cantilever allowing light transmission for magneto-optical experiments like BLS. The sample and the antenna device can be moved independently from each other. (b) Side view. (c) An SMA adapter connects the suspended antenna to external microwave equipment. The suspended antenna is fabricated on a 100 micrometer thick glass substrate which is bonded and fixed on a coplanar waveguide connected to the SMA adapter. A current through the short circuits between ground and signal lines generates a dynamic magnetic Oersted field which can be used to excite the magnetization. (d) Straight line short circuit (e) Omega loop short circuit for effective out-of-plane excitation. }
\end{center}
\end{figure}

The investigation of magnetization dynamics for research on new computational techniques that could allow for a faster data processing with higher energy efficiency is of high interest. 
Magnonics and spintronics are promising fields in order to achieve this revolution in data processing by using spin wave-based computing\cite{K. Vogt, J.Cramer}.  
However, for these new techniques more fundamental and material research needs to be done. Especially novel materials with low damping for improved spin wave propagation or enhanced nonlinear effects at moderate pump powers as well as excitation methods are needed. 

In order to compete with todays computational schemes, the development of new techniques needs to be nanoscalable and makes thin film technology necessary. Therefore, new materials are mostly deposited as thin films firstly which allows for the characterization by standard techniques like FMR, MOKE\cite{Bader} and vibrating sample magnetometry (VSM)\cite{Foner}.

On the other hand, to investigate spin wave propagation in materials and microstructures other techniques have to be used. They can be measured electrically by propagating spin wave spectroscopy \cite{V. Vlaminck, Grundler, Lucassen} 
 using antennas with a defined wave vector (k) excitation but without spatial resolution of the spin wave propagation. Another promising technique is Brillouin light scattering microscopy (${\mu}$BLS)\cite{Sebastian2015}, which is able to detect coherently excited spin waves as well as incoherent thermal spin waves spatially and time resolved. Moreover, the spin wave dispersion relation can be investigated by phase resolved measurements. This technique is based on the inelastic scattering of light by spin waves, which leads to a frequency shift of the scattered laser light. 
Regardless of whether electrical or optical techniques are used, antennas for the excitation of the spin waves need to be patterned on the sample
, which is a rather complex, time and resource consuming process. Moreover, on conductive substrates (e.g. epitaxial iron grown on single-crystal copper)\cite{L. Flajsman} where an insulating layer below the antenna is necessary, contacting them with probes can destroy this insulation. 
Here we demostrate an alternative approach. We separate the antenna used for excitation of the magnetization from the sample. 
This makes the antenna freely-positionable with respect to the sample under investigation and allows for probing the spin wave dynamics locally whereas standard FMR measurements integrate over a large area\cite{H. K. Lee, I.Barsukov}. Furthermore, the relative orientation between the antenna and the magnetic element can be easily varied\cite{R. Meckenstocka}. This allows to probe the spin waves in the same magnetic microstructure for different angles of the applied magnetic field. Since the antenna can be easily removed in contrast to antennas which are patterned directly on the sample, 
different dynamic field configurations can be applied. 
The possibility of reusing the antennas 
offers the additional advantage that these single antennas need to be characterized only once regarding their frequency dependent transmission function. 


\subsection*{Specifications of the antenna device}

Fig.~\ref{fig7} (a) shows schematically the suspended antenna in operation. It is separated from the sample under investigation and patterned on a 
glass substrate which is held by an adapter unit. It can be moved along all three axes in order to reach every measurement position on the sample, which is given by the focused laser spot in magneto-optical experiments like ${\mu}$BLS. Fig.~\ref{fig7} (b) and (c) present the actual device in side view and top view, respectively. A subminiature A (SMA) adapter at the bottom is used to connect the suspended antenna to signal generators by standard SMA cables. This adapter is soldered to a coplanar waveguide fabricated on a high frequency printed circuit board (PCB). On this PCB the glass substrate is fixed and wirebonded using 20 ${\mu}$m x 250 ${\mu}$m ribbons. It is the actual part carrying the antenna structure, which was patterned by EBL. The conductive layer 
 was deposited by electron beam evaporation as well as an additional \ce{SiO2} insulating layer.
This layer protects the antenna's coplanar waveguide to be shorted by a conductive sample if they are in contact. 
The signal and both ground lines are short circuited at the end in order to allow a current flow generating a dynamic magnetic Oersted field that excites the magnetization in the sample. The shape of the short-circuit can be adjusted regarding the experimental needs for example straight line-wires as shown in Fig.~\ref{fig7} (d) and (e) for in-plane or omega-shaped loops for out-of-plane excitation, respectively.

\begin{figure}[b]
\begin{center}
\scalebox{1}{\includegraphics[width=8.5 cm, clip]{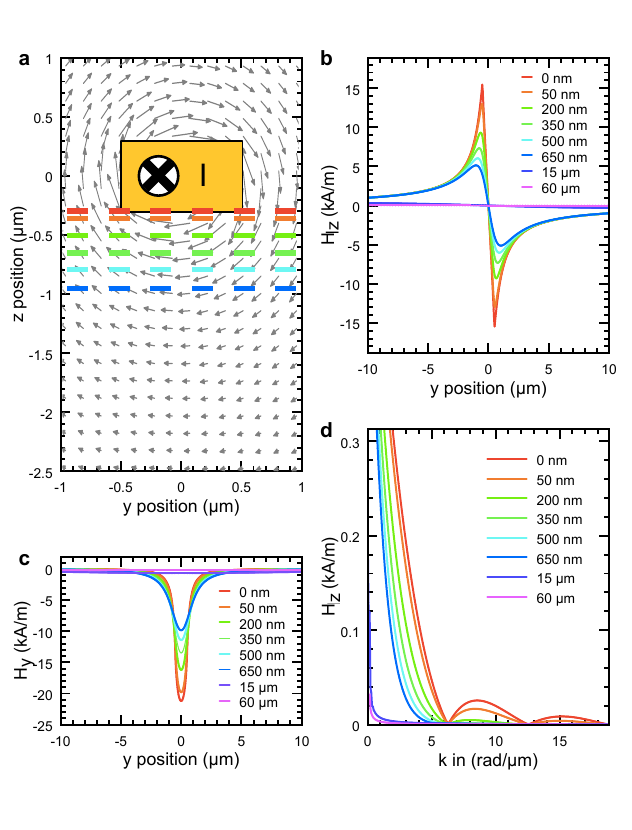}}
\caption{\label{fig6}(color online) (a) Calculated Oersted field around the short circuit of the antenna with rectangular shape (parameters: width: 1 ${\mu}$m, height: 600 nm, dc current: 63 mA). The strength and local orientation of the field are denoted by the arows. The dashed lines indicate the z position where the field distribution was analyzed. 
(b) z component of the Oersted field extracted along the dashed lines for different distances between antenna and sample. (c) y component of the Oersted field extracted along the dashed lines for different distances between antenna and sample. 
(d) Fourier transform of the field distributions shown in (b). 
}
\end{center}
\end{figure}

An important property of the antenna is the possibility to change the distance between sample and antenna. Fig.~\ref{fig6} (a) shows the calculated Oersted field for a current carrying wire with a rectangular shape. 
The width is 1 micrometer and the height 600 nm corresponding to the actual dimensions of the used antenna short circuit. Since the skin depth of a microwave current in a frequency range up to 20 GHz is larger then the antenna dimensions, the field calculations can be approximated with static currents and a value of 63 mA was chosen. It corresponds to a microwave power of 20 dBm assuming that the wire is a 50 Ohm termination of a microwave waveguide. 
If the sample is in direct contact with the antenna, the magnetic field interacting with the sample is extracted along the red dashed line. The out-of-plane (z) and in-plane (y) component of this field were extracted and are shown in Fig.~\ref{fig6} (b) and Fig.~\ref{fig6} (c) as the 0 nm case, respectively. 
Moreover, the plots show the field distribution for a variable distance between antenna and sample from 0 to 60 ${\mu}$m. By increasing the distance, the maximal field strength decreases, but its homogeneity increases. To investigate this effect further, we made a Fourier transform of the spatial distribution of the magnetic field in order to obtain information about spatial frequencies, which are 
plotted in Fig.~\ref{fig6}(d). 
It can be clearly seen, that the field amplitudes for small k vectors are significantly larger than for large k vectors. By increasing the distance between antenna and sample, the field amplitudes at higher k values are even more suppressed. Only for larger distances the field amplitudes drop dramatically. 


\subsection*{Time-efficient characterization of magnetic films}

\begin{figure}[b]
\begin{center}
\scalebox{1}{\includegraphics[width=8.5 cm, clip]{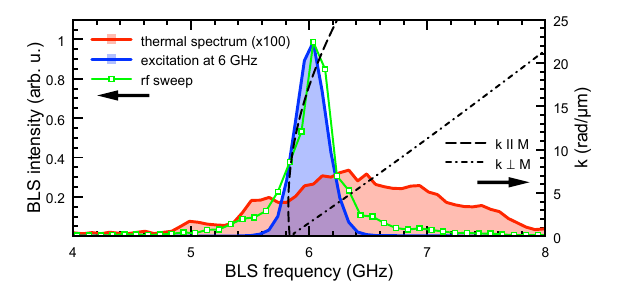}}
\caption{\label{fig3}(color online) Comparison of BLS spectra without (red) and with (blue) antenna excitation at 6 GHz and a nominal power of 30 dBm measured on a Pt(4.5 nm)/Cu(0.87 nm)/Py(5 nm)/Al(3.5 nm) 
film. A thermal spin wave band can be measured between 5 and 8 GHz at ${B}_\mathrm{ex}=51\,$ mT. By applying a microwave current to the antenna the signal at the exciation frequency can be increased by a factor of about 400. The green curve shows the reached intensity maxima during the rf excitation at the corresponding frequencies. The dashed and dash-dotted line show the spin wave band between propagating spin waves parallel or perpendicular to the magnetization direction for the used parameters: ${M}_\mathrm{S}=640\,$ kA/m, ${\gamma}_{}/2{\pi}=28\,$ GHz/T and ${A}_\mathrm{ex}=10\,$ pJ/m.}
\end{center}
\end{figure}

\begin{figure}[b]
\begin{center}
\scalebox{1}{\includegraphics[width=8.5 cm, clip]{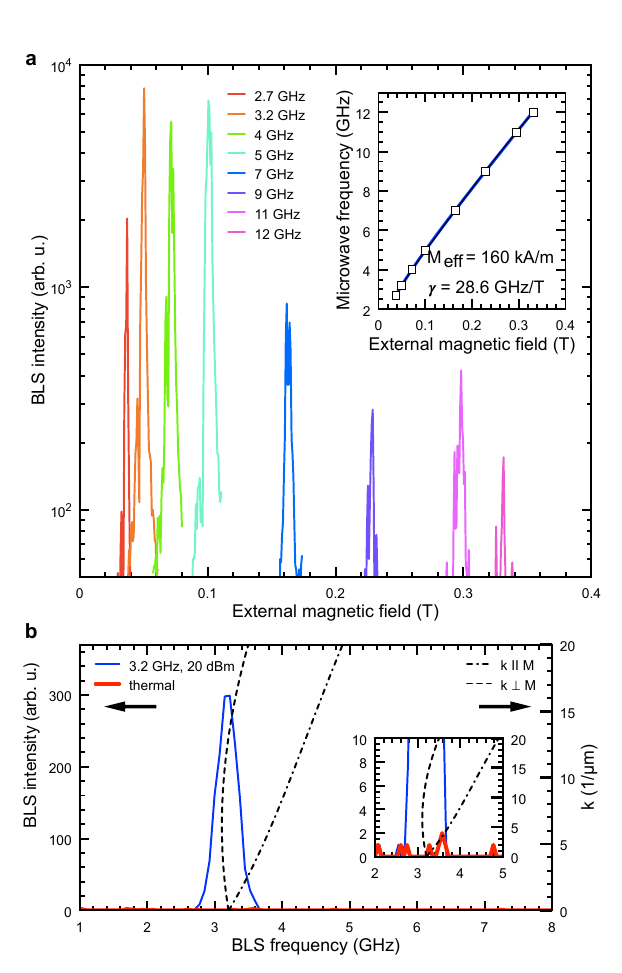}}
\caption{\label{fig4}(color online) (a) Ferromagnetic resonance peaks of a 50 nm YIG film measured by BLS during fieldsweeps with a stepsize of 0.7 mT using fixed microwave frequencies of the antenna. The inset shows the extracted resonance fields and corresponding excitation frequencies. The data was fitted by the Kittel formula in order to obtain ${M}_\mathrm{eff}=160\,$ kA/m and ${\gamma}_{}/2{\pi}=28.6\,$GHz/T. (b) Direct comparison of the BLS intensity with and without antenna excitation. The dashed and dashed-dotted line indicate the borders of the spin wave band calculated with these parameters: ${M}_\mathrm{S}=160\,$ kA/m, ${\gamma}_{}/2{\pi}=28.6\,$ GHz/T and ${A}_\mathrm{ex}=5\,$ pJ/m. 
The inset shows the enlarged area at the excitation frequency. No signal could be obtained without antenna excitation.}
\end{center}
\end{figure}

\begin{figure}[b]
\begin{center}
\scalebox{1}{\includegraphics[width=8.5 cm, clip]{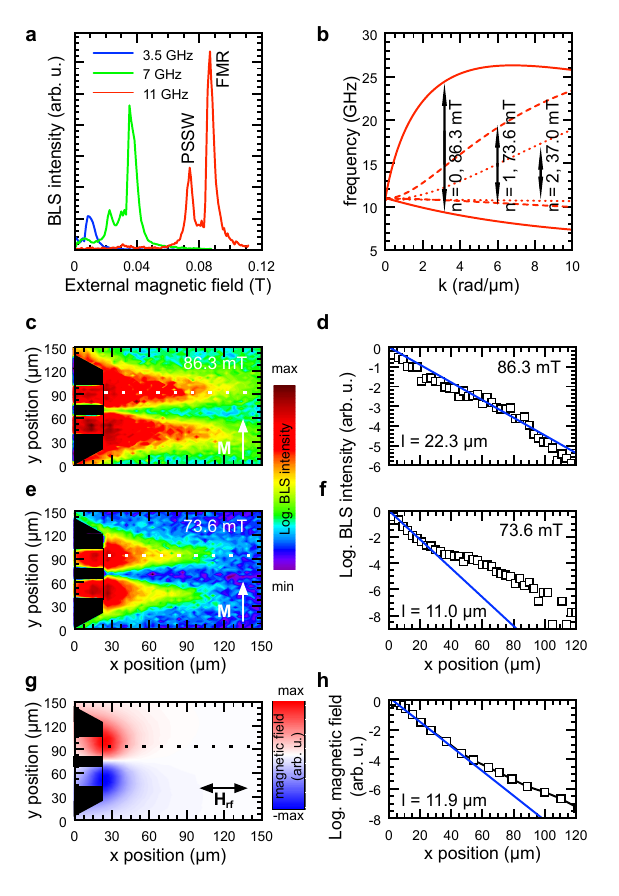}}
\caption{\label{fig5}(color online) (a) Ferromagnetic resonance peaks measured by BLS during field sweeps at fixed microwave frequencies at the antenna. (b) Spin wave dispersion relation calculated for a 240 nm CoFeB film and ${\mu}_\mathrm{0}{M}_\mathrm{S}=1.7\,$ T, ${A}_\mathrm{ex}=46\,$ pJ/ m for different external fields and quantizations n over the film thickness. The red (red-dashed) curves show the dispersion relation of the first n=0 (second, n=1) order of quantization over the film thickness. (c),(e) Spatially resolved measurements of the spin wave intensity around the antenna (drawn in black) at an external field of 86.3 mT and 73.6 mT, respectively. (d),(f) Extracted BLS intensity along the white dotted line in (c) and (d) for the determination of the decay length for the external magnetic field of 86.3 mT and 73.6 mT, respectively. (g) Calculated magnetic in-plane field distribution around the antenna. (h) Exctracted field distribution along the black dotted line in (g).}
\end{center}
\end{figure}

The magneto-dynamic properties of newly developed materials can effectively be characterized by BLS in the case of sufficient cross section of the light with the material, which can therefore probe the dynamic magnetic modes. Usually the new materials are deposited as thin films or even as nanoparticles on flat substrates. The typical approach for first tests is the measurement of the thermally excited, incoherent spin waves in these samples, which can, depending on the material, become a very time-consuming and rather inefficient task. 
In some materials, which are even transparent for the used laser light, these measurements can take hours or the obtained signals are even hidden in the counting noise of the photo detector.
In this section we demonstrate a much more efficient approach for the characterization of materials and the possibility to get an estimate for material parameters by carrying out optical FMR measurements. 
As a first proof of concept 
 we show in Fig.~\ref{fig3} the direct comparison of the measured BLS counts of thermally excited spin waves in a Pt(4.5 nm)/Cu(0.87 nm)/Py(5 nm)/Al(3.5 nm) 
 film with the strong signal when the antenna is used to excite the magnetization dynamics. An external inplane field of 51 mT was applied and the antenna was connected to a signal generator outputting 6 GHz at a nominal power of 30 dBm. Notice, that both measurements were performed through the glass cantilever in order to get comparable measurement conditions. 
The direct comparison shows an increase of the BLS counts by a factor of about 400 for the same measurement time. 
Additionally, we plotted the peak intensity for different microwave excitations to demonstrate that dynamics can be driven over a wide frequency range. Even though the largest intensity is reached around the ferromagnetic resonance we observe a significant increase for higher frequencies up to 6.6 GHz, which corresponds to a wave vector of 7 rad/${\mu}$m. That is a clear demonstration of the utility of our novel technique in order to reduce measurement time. Notice, that we used a sample that shows a sufficient thermal spectrum within several minutes in order to compare the BLS counts with and without antenna excitation. In addition we calculated the spin wave dispersion relation for a 5 nm thick Py thin film to compare it with the frequency range of the measured thermal spin wave band and the frequency with the highest spin wave amplification caused by the antenna. Regarding the calculations shown in Fig.~\ref{fig6} we expect an efficient coupling of the microwave field of the antenna to the magnetization close to the FMR mode. This could be verified since the highest amplification of the spin waves was found at 6 GHz. Please note, that the reduction of spin wave intensities above 7.5 GHz is due to experimental limitations. Here the used microscope objective limits the detection to k-vectors below 20 rad/${\mu}$m\cite{Sebastian2015}.

As a next step, we show an optical FMR measurement using BLS on a 50 nm thick yttrium iron garnet (\ce{Y3Fe5O12}, YIG) film, which was grown epitaxially by pulsed laser deposition on a \ce{Gd3Ga5O12}(111) substrate. YIG is transparent for the used 532 nm laser and backscattering due to thermal spin waves has a very small cross-section. 

However, by using the antenna for the excitation of the magnetization the measurement of ferromagnetic resonance curves by sweeping the external in-plane field with a stepsize of 0.7 mT at fixed frequencies becomes an efficient way to characterize the local material properties. In Fig.~\ref{fig4}(a) the resonance peaks are shown for different applied frequencies to the antenna at a nominal power of 20 dBm. The resonance peaks were fitted using Lorentz functions in order to get the resonance fields. These values are plotted in the inset together with the used frequencies and fitted by the Kittel formula. The extracted values for the effective magnetization are ${M}_\mathrm{eff}=160\,$ kA/m and for the gyromagnetic ratio ${\gamma}_{}/2{\pi}=28.6\,$GHz/T. 
Additionally, we show in Fig.~\ref{fig4}(b) a direct comparison of the BLS intensity with and without antenna excitation of the sample at 3.2 GHz (compare to Fig.~\ref{fig4}(a)). 
The dashed and dash-dotted lines indicate the borders of the spin wave band. The highest spin wave intensity can be excited around the FMR frequency. The inset shows the enlarged area of the excitation in order to show the large difference between the measured spin wave intensity with and without the antenna excitation. No signal can be obtained, if the antenna does not excite the YIG thin film. This strongly confirms the utility of our new technique since it opens the possibility to investigate materials which otherwise can not be measured with spatial resolution in a reasonable time by thermal excitation only.

As an additional example, we show the ferromagnetic resonance curves measured on a 240 nm thick \ce{Co40Fe40B20} sample. Fig.~\ref{fig5}(a) shows resonance curves for 3 different applied frequencies at a nominal power of 20 dBm and a distance of 15 micrometers between antenna and sample.
The resonance curve obtained for the 11 GHz excitation will be of interest in the following part. It is characterized by two resonances at different in-plane field values of 73.6 mT and 86.3 mT. By the calculation of the spin wave dispersion relation 
 using ${A}_\mathrm{ex}=46\,$ pJ/ m, ${\mu}_\mathrm{0}{M}_\mathrm{S}=1.7\,$ T and 
the given externernal fields at the resonance, it was found that the one at the higher field can be attributed to the FMR (n=0) and the resonance at the lower field to the perpendicular standing spin wave (PSSW) mode (n=1) (compare to Fig.~\ref{fig5}(b)). It shows, that the next quantization order n=1 of the spin waves over the film thickness is located at 11 GHz for ${\mu}_\mathrm{0}{H}_\mathrm{ext}=73.6\,$ mT. Therefore, two resonances at close magnetic fields are expected, if the excitation frequency is kept constant as in this experiment. 
We also calculated the spin wave band for the n=2 quantization over the film thickness and found an expected resonance at 11 GHz for ${\mu}_\mathrm{0}{H}_\mathrm{ext}=37.0\,$ mT. This could be an indication that the little increase of intensity at ${\mu}_\mathrm{0}{H}_\mathrm{ext}=37.0\,$ mT in the measurement shown in Fig.~\ref{fig5}(a) is the excitation of spin waves with this particular quantization.
For the n=0 and n=1 mode we did spatially resolved BLS measurements to have a deeper insight in the excitation profile of the antenna on the film. Figs.~\ref{fig5}(c) and (e) show the results for 86.3 mT and 73.6 mT, respectively. The position of the antenna is plotted in black. It is cleary visible, that both short circuits of the coplanar wave guide excite the magnetization equally indicating that the same amount of current is flowing through both of them. Starting from the short circuits two narrow strong excitation regions are detected. Between them the intensity drops significantly. This can be attributed to the 180${^\circ}_{}$ phase shift between both excitations. While the excited area for the measurement at 86.3 mT seems to be strongly expanded, the one at 73.6 mT seems to be located closer to the antenna. To investigate this further, we extracted the BLS intensity along the white dotted lines and plotted them in Fig.~\ref{fig5}(d) and (f), respectively. An exponential fit for the measurement at the higher field shows a close to linear behavior and allows for the determination of a decay length of 22.3 micrometers for the BLS intensity, indicating a propagating spin wave mode. For the smaller field, the extraction shows a deviation from a clear linear behavior, indicating a localized, directly excited mode in the close vicinity of the antenna. We fitted only the points close to the antenna in order to obtain a decay length for comparison. A value of 11.0 micrometers was obtained. From the fits it can be confirmed, that the excitated mode at 86.3 mT extends further. The different propagation range of both modes can be explained by the spin wave dispersion relation. The n=0 mode includes much larger slopes i.e. group velocities. Therefore, an extended propagation of these spin waves is expected compared to the spin waves with n=1, which have smaller slopes. Since the area of excitation seemed to be quite large for even ${\mu}_\mathrm{0}{H}_\mathrm{ext}=73.6\,$ mT, we simulated the magnetic field distribution around the antenna with COMSOL for the given antenna distance to the sample and plotted the in-plane component in Fig.~\ref{fig5}(g). It can be clearly seen, that the magnetic field generated on both sides has a phase difference of 180${^\circ}_{}$ generating a point of zero magnetic field in between. Similar to the measurements we extracted the field distribution along the dotted line and plotted it in Fig.~\ref{fig5}(h). It is in an qualitative agreement with Fig.~\ref{fig5}(f) showing no linear behavior in the logarithmically scaled plot. The fit of values close to the antenna reveal a decay length of about 11.9 micrometer, which is in good agreement with the value obtained from Fig.~\ref{fig5}(f). Therefore we conclude that the n=1 mode at ${\mu}_\mathrm{0}{H}_\mathrm{ext}=73.6\,$ mT is a direct excitation in the close vicinity of the antenna rather than a propagating spin wave. 

\begin{figure*}
\begin{center}
\scalebox{1}{\includegraphics[width=17 cm]{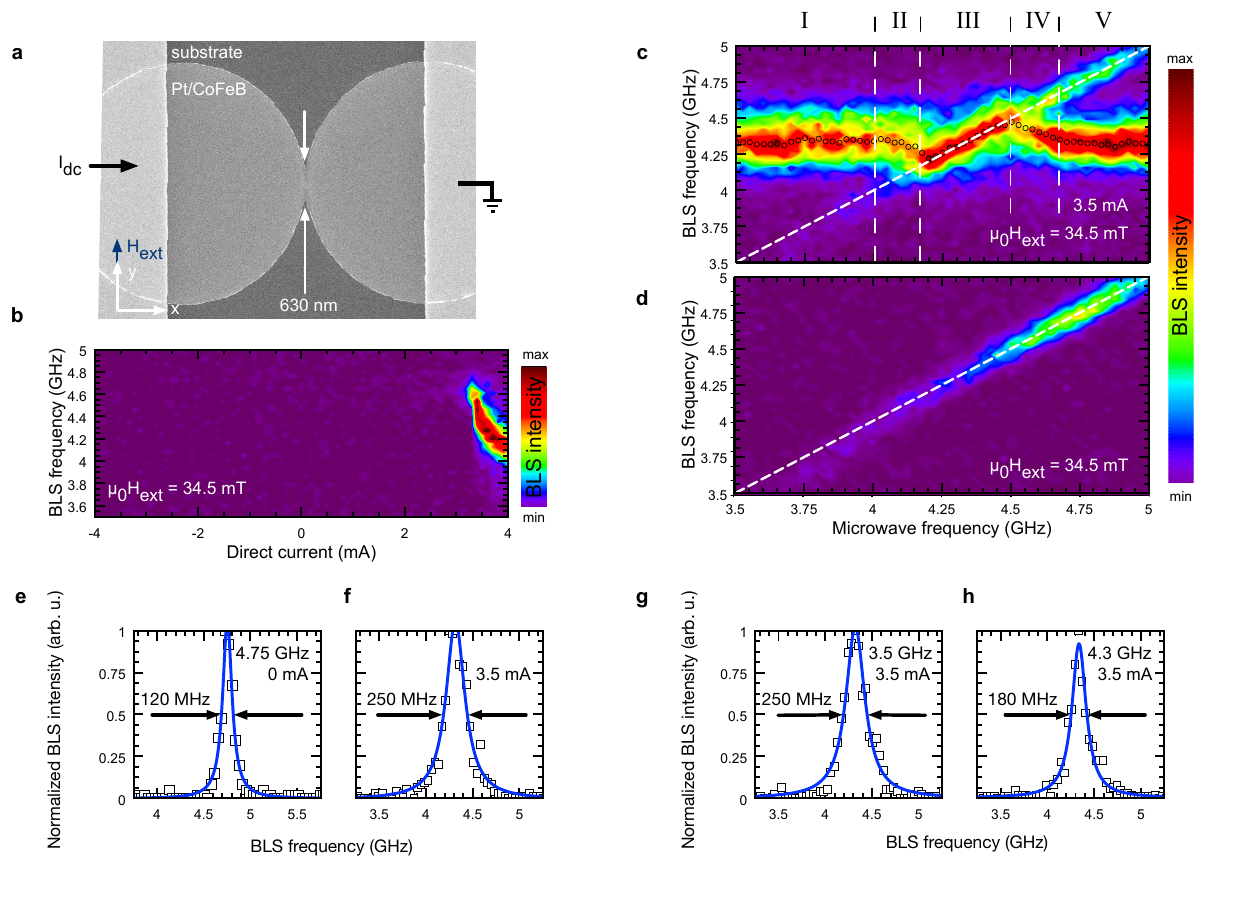}}
\caption{\label{fig1}(color online) (a) SEM image of the SHNO with a 630\,nm wide constriction. The external magnetic field $\mu_\mathrm{0}H_\mathrm{ext}$ was applied perpendicular to the dc current. (b) BLS intensity detected between 3.5 and 5\,GHz as a function of the dc current ranging from $-4.0$ to $+4.0$\,mA. A magnetic field of $\mu_\mathrm{0}H_\mathrm{ext}=34.5\,\mathrm{mT}$ was applied. 
For currents above 3.2\,mA, a rapid increase in the intensity is observed, marking the threshold for the excitation of auto-oscillations. 
(c) Locking characteristics of the SHNO in Fig.~\ref{fig1} (a) measured as a function of microwave frequency  for a fixed $I_\mathrm{dc}=3.5\,\mathrm{mA}$ and $\mu_\mathrm{0}H_\mathrm{ext}=34.5\,\mathrm{mT}$. The detected intensity is colour coded with violet (red) representing minimum (maximum) values. Black dots mark the frequencies $f_\textrm{ao}$  of the excited auto-oscillations. Five different operating regimes can be identified: I,V free-running state of the auto-oscillation, II, IV frequency down- and up-pulling, respectively, and III the locked state. (d) Reference measurement without an applied dc current through the SHNO. (e) Linewidth of the microwave excitation at 4.75 GHz and a nominal power of 16 dBm. (f) Linewidth of the auto-oscillations at a dc current of 3.5 mA. (g) Linewidth of the auto-oscillations outside the locking state at 3.5 mA and an external stimulus of the antenna at 3.5 GHz. (h) Linewidth of the auto-oscillations in the locked state at 3.5 mA and an external stimulus of 4.3 GHz.
}
\end{center}
\end{figure*}

\subsection*{Injection-locking of spin Hall nano-oscillators}

In this section we demonstrate the possibility to use the antenna device to do injection-locking experiments with spin Hall nano-oscillators (SHNO). While in recent publications\cite{T.Hache, Spicer2018, K.Wagner} microwave currents are directly applied to the SHNO to create a torque on the magnetization we demonstrate here the synchronization of the auto-oscillations of the magnetization to the dynamic Oersted field of the suspended antenna device. The used SHNO (Fig.~\ref{fig1}(a)) was fabricated using EBL and magnetron sputtering. The layer stack consists of Ta(2 nm)/Pt(7 nm)/\ce{Co40Fe40B20}(5 nm) /Ta(2 nm). The Au contacts were deposited by thermal evaporation. Ta was used as seed and capping layer.


The applied dc current is mostly flowing through the Pt  and generates a pure spin current via the spin Hall effect\cite{Hirsch1999,Ando2008, Hoffmann2013}. This pure spin current enters through the interface into the magnetic  \ce{Co40Fe40B20}  layer. The magnetization of this layer is aligned perpendicular to the dc current by an external magnetic field, which is a necessary condition to get auto-oscillations of the magnetization.\cite{Slonczewski,Berger,Slavin,Demidov2014,Demidov2014locking,Demidov2012,Yang2015,Duan2014,Kiselev} The auto-oscillations are localized in the 630 nm wide constriction, where the highest current density is reached.\cite{Dvornik2018} Fig.~\ref{fig1}(b) shows the intensity of the auto-oscillations measured by BLS in dependence of the applied current at an external magnetic field of 34.5 mT. The auto-oscillations occur only in one current direction whereas in the opposite current direction an increased damping of the magnetization is forced. Please note also the negative nonlinear frequency shift of the auto-oscillations during the increase of current and the onset of the auto-oscillations, which is a typical property for SHNO with inplane magnetization.\cite{Slavin}
The increase of intensity and the negative nonlinear frequency shift correlate with each other.

After the demonstration of auto-oscillations in the SHNO, the antenna was positioned close to the constriction and connected to a signal generator. The microwave frequency through the antenna was swept between 3.5 and 5 GHz in 25 MHz steps at a nominal power of 16 dBm and a dynamic Oersted field was generated affecting the magnetic layer of the SHNO. A dc current of 3.5 mA was applied to the SHNO at an external magnetic field of 34.5 mT. 
In Figure~\ref{fig1}(c) it is demonstrated that the auto-oscillations can synchronize to the dynamic Oersted field in a certain locking range. We separate the locking process into five segments\cite{T.Hache}. If the difference between the frequency of the external stimulus and the auto-oscillations is too large then there is no synchronization possible and the auto-oscillation frequency stays constant (segment I and V). However, if the difference between microwave and auto-oscillation frequency is reduced a frequency pulling sets in. The auto-oscillations frequency is stepwise attracted to the frequency of the external stimulus (segment II and IV). Within segment III the auto-oscillations are synchronized to the dynamic antenna field. This is indicated by the linear correlation between the external driven microwave frequency through the antenna and the auto-oscillation frequency in this segment. The included circles are obtained by fitting the measured spectra by Lorentz functions. Figure~\ref{fig1}(d) shows the excitation from the antenna without auto-oscillations meaning, that the dc current was switched off during this measurement. It can be clearly seen, that the direct spin wave excitation within the locking range is much weaker than the auto-oscillations intensity. Please note, that both plots are normalized to each other. Moreover, it can be seen from Figure~\ref{fig1}(d) that the intensity maximum is reached around 4.75 GHz marking the FMR. From this it can be concluded, that the auto-oscillations are located at lower frequencies compared to the FMR. This is a well-known property of SHNO with inplane magnetization and fits to the behaviour of Figure~\ref{fig1}(b), where a red shift was detected. 
Figure~\ref{fig1}(e)-(h) compares the properties of the excited spin wave modes within the SHNO. Figure~\ref{fig1}(e) shows an extracted BLS spectrum with an antenna excitation at 4.75 GHz and without dc current. 
We fitted it with a Lorentzian peak and obtained a linewidth value of 120 MHz, which can be treated as the linewidth broadening of the BLS setup itself. Figure~\ref{fig1}(f) shows in comparison a BLS spectrum, where only a dc current of 3.5 mA was present with no microwave current applied to the antenna. It can clearly be seen, that the coherence of the auto-oscillations is much smaller since a a linewidth of 250 MHz was obtained. We directly compare this value with the linewidth of the auto-oscillations where the additional stimulus by the antenna is applied. In Figure~\ref{fig1}(g) we plot the linewidth of the auto-oscillations at a dc current of 3.5 mA and an applied microwave frequency of 3.5 GHz at a nominal power of 16 dBm through the antenna. Since the applied microwave frequency is very different from the auto-oscillation frequency (about 4.3 GHz), no interaction is detected which is confirmed by the same linewidth of the auto-oscillations of about 250 MHz. However, the linewidth significantly changes if a microwave frequency of about 4.3 GHz is applied to the antenna at the same nominal power. Figure~\ref{fig1}(h) shows the reduction of the linewidth to 180 MHz in the locked state. It can be clearly seen, that the injection locking of the auto-oscillations results in an increased coherence.

\subsection*{Conclusion}

We have demonstrated the potential of a suspended antenna by comparing the obtained signal of thermal spin wave measurements with the 400 times larger signal by additional excitation. We have shown how  thin films can be investigated with high spatial resolution by doing optical FMR measurements. Due to the additional excitation the measurements on samples with very low signal such as YIG thin films become not only possible but also efficient. We have shown the excitation of propagating spin waves in 240 nm thick CoFeB films and explained the excitation characteristics by comparison to a simulation of the antenna field. Moreover, we were able to synchronize spin Hall nano-oscillators to the dynamic Oersted field generated by the antenna. Behind the scope of this work the application of localized microwave fields is also interesting for the investigation of magnetic nanoparticles and the excitation in optically detected magnetic resonance (ODMR) experiments, e.g. of SiC vacancy centers\cite{Astakhov}. Additionally, the antenna device could be used also in non-optical experiments for excitation of the magnetization wherby the detection is realized by an vector network analyzer.

\subsection*{Acknowledgments}
Financial support by the Deutsche Forschungsgemeinschaft is gratefully acknowledged within program SCHU2922/1-1.  
The antennas were fabricated at the CEITEC Nano Research Infrastructure (ID LM2015041, MEYS CR, 2016-2019). We thank B. Scheumann for deposition of the Au films. Lithography was done at the Nanofabrication Facilities (NanoFaRo) at the Institute of Ion Beam Physics and Materials Research at HZDR. I.B. acknowledges support by the National Science Foundation under grant No. ECCS-1810541.


\end{document}